\documentclass[twocolumn]{aastex63}
\usepackage{lineno}
\shorttitle{Are CL-NLS1s at a Particular Evolutionary Stage?}
\shortauthors{Wang et al.}

\begin{document}

\title{Are ``Changing-Look'' Active Galactic Nuclei Special in the Coevolution of 
Supermassive Black Holes and their Hosts? II. The Case of Changing-Look Narrow-Line Seyfert 1 Galaxies}

\correspondingauthor{J. Wang \& D. W. Xu}
\email{wj@nao.cas.cn, dwxu@nao.cas.cn}

\author{J. Wang}
\affiliation{National Astronomical Observatories, Chinese Academy of Sciences, Beijing 100101, People's Republic of China}

\author{S. Jin}
\affiliation{Guangxi Key Laboratory for Relativistic Astrophysics, School of Physical Science and Technology, Guangxi University, Nanning 530004,
People's Republic of China}
\affiliation{National Astronomical Observatories, Chinese Academy of Sciences, Beijing 100101, People's Republic of China}
\affiliation{School of Astronomy and Space Science, University of Chinese Academy of Sciences, Beijing, People's Republic of China}

\author{D. W. Xu}
\affiliation{National Astronomical Observatories, Chinese Academy of Sciences, Beijing 100101, People's Republic of China}
\affiliation{School of Astronomy and Space Science, University of Chinese Academy of Sciences, Beijing, People's Republic of China}

\author{WeiKang Zheng}
\affiliation{Department of Astronomy, University of California, Berkeley, CA 94720-3411, USA}

\author{Thomas G. Brink}
\affiliation{Department of Astronomy, University of California, Berkeley, CA 94720-3411, USA}

\author{S. Komossa}
\affiliation{Max-Planck-Institut für Radioastronomie, Auf dem Hügel 69, 53121 Bonn, Germany}
\affiliation{National Astronomical Observatories, Chinese Academy of Sciences, Beijing 100101, People's Republic of China}

\author{Alexei V. Filippenko}
\affiliation{Department of Astronomy, University of California, Berkeley, CA 94720-3411, USA}

\author{J. Y. Wei}
\affiliation{National Astronomical Observatories, Chinese Academy of Sciences, Beijing 100101, People's Republic of China}
\affiliation{School of Astronomy and Space Science, University of Chinese Academy of Sciences, Beijing, People's Republic of China}






\begin{abstract} 

The evolutionary role of the so-called ``changing-look'' (CL) active galactic nucleus (AGN), which is characterized by spectral-type transitions 
within $\sim10$~yr, has been suggested in the past few years. 
By focusing on CL-AGNs having spectra similar to those of broad-line Seyfert 1 galaxies, some authors have proposed that CL-AGNs tend to be at a special
evolutionary stage associated with intermediate-to-old stellar populations.
Here we attempt to verify this evolutionary role by extending the sample to CL narrow-line 
Seyfert 1 (NLS1) galaxies, which are believed to be ``young'' AGNs with a less massive supermassive black hole and high accretion rate. Combining the 
recent large NLS1 catalog provided by Paliya et al. (2024) and the SDSS-V DR19 spectral survey returns only three CL-NLS1s out of a parent sample of 884 objects, reinforcing the rarity of CL-NLS1s. Subsequent spectral analysis shows that the evolutionary role mentioned above still holds, although CL-NLS1s tend to
occupy the young end of the intermediate-old population. Finally, we propose that  
off-center SDSS spectra caused by the ``fiber drop'' effect have great potential for  determining the properties of the narrow-line region of NLS1s.
\end{abstract}



\section{Introduction} \label{sec:intro}

As a special kind of active galactic nuclei (AGNs), ``changing-look'' AGNs (CL-AGNs)
manifest themselves in a spectral transition between Type~1, intermediate type, and Type~2 within a timescale of years to decades, due to  
a temporary appearance or disappearance of their broad emission lines 
(see reviews by Ricci \& Trakhtenbrot 2022; Komossa \& Grupe 2023, 2024; Komossa et al. 2024).
So far, $\sim1000$ CL-AGNs have been identified in the past decade,
thanks to great developments in the area of multi-epoch spectroscopy
(e.g., MacLeod et al. 2019; Sheng et al. 2017, 2020; 
Wang et al. 2018, 2019, 2023; Yang et al. 2018, 2025; Guo et al. 2019; Trakhtenbrot et al. 2019; Yan et al. 2019; Hon et al. 2022; 
Parker et al. 2019; L{\'o}pez-Navas et al. 2023; Guo et al. 2024, 2025; Wang et al. 2025; Zeltyn et al. 2024; Dong et al. 2025).

There is accumulating evidence supporting the hypothesis that 
the CL phenomenon is caused by variations in the accretion rate of
a supermassive black hole (SMBH; e.g., Sheng et al. 2017,
2020; Yang et al. 2018; MacLeod et al. 2019; Feng et al. 2021; Shen et al. 2025; Lu et al. 2025). Other possible origins include, for instance, an accelerating outflow
(e.g., Shapovalova et al. 2010), a variation of the obscuration (e.g., Elitzur 2012),
a tidal torque on the mini-disk of each SMBH in a close binary of SMBHs with a high eccentricity (Wang \& Bon 2020). Tidal disruption events (TDEs) can mimic CL-AGNs in rare cases (e.g., Merloni et al. 2015;
Blanchard et al. 2017; Padmanabhan \& Loeb 2021; Wang et al. 2024b; Komossa \& Grupe 2024), and 
it has been speculated that individual AGN like 1ES~1927+654 was powered by a TDE 
(e.g., Li et al. 2024; Cao \& You 2024). 
However, TDEs are very rare events and are most reliably identified in quiescent host galaxies 
(Komossa \& Bade 1999) without long-lived accretion disk. In order to identify TDEs in AGNs,
mutliple criteria have to be considered (see the detailed discussion by
Komossa \& Grupe 2024). For instance, TDEs typically show decay of $\propto t^{-5/3}$ that is
inconsistent with those of CL-AGNs (e.g., Yao et al. 2025).

The CL phenomenon challenges steady-state accretion disk models (e.g., Shakura \& Sunyaev 1973). Cases of 
exceptionally high amplitude of variability in the optical band, and the 
identification of rapid CL events in luminous quasars
(e.g., La Massa et al. 2015, Wang et al. 2018),
have let Lawrence (2018) to emphasize the ``viscosity crisis,'' as these 
systems vary much faster than timescales in steady-state disks. Some 
revisions of accretion-disk models have been proposed 
to alleviate the crisis by considering a local disk thermal instability 
(e.g., Husemann et al. 2016), a Lightman-Eardley radiation-pressure instability 
(Grupe et al. 2015), a magnetic field (e.g., Ross et al. 2018;
Stern et al. 2018; Dexter \& Begelman 2019; Feng et al. 2021;
Pan et al. 2021; Cao et al. 2023), or a transition from advection-dominated accretion 
flow (ADAF) to the Shakura–Sunyaev disk (Shakura \& Sunyaev 1973) at the inner accretion disk (Li \& Cao 2025).

By focusing on the host galaxies of CL-AGNs with intermediate spectral types,
Wang et al. (2023) proposed that CL-AGNs tend to be associated with an intermediate-old 
($\sim10^9$~yr) stellar population. This population correspond to a specific evolutionary
stage being from ``feast'' to ``famine'' fueling of the SMBH. 
The intermediate-old stellar population is supported by Liu et al. (2021) and Jin et al. (2022) 
for broad-line Seyfert 1 galaxies and quasars.  
This difference in the properties of host galaxies has, however, been recently disputed by Verrico et al. (2025), who suggests
that CL-AGNs and non-CL-AGNs tend to be of similar star-formation properties by modeling the 
the stellar populations by the Prospector software (e.g., Johnson et al. 2021).

One should bear in mind that,  
in all the identified CL-AGNs, the objects with a narrow-line Seyfert 1 nucleus (NLS1-like\footnote{NLS1s are defined as broad-line AGNs with relatively narrow broad Balmer
emission (full width at half-maximum intensity $\mathrm{FWHM(H\beta)\lesssim 2000\ km\ s^{-1}}$), weak [\ion{O}{3}] line 
emission ($\mathrm{[O~III]/H\beta}<3$), strong \ion{Fe}{2} complex emission,  
and often supersoft X-ray spectra (see Komossa 2008 for a review).})
spectrum (hereafter CL-NLS1s) are still rare at the current stage. 
To our knowledge, there are only in total seven identified CL-NLS1s with relatively narrow
Balmer broad emission lines (i.e., $\mathrm{FWHM(H\alpha)}$ or 
$\mathrm{FWHM(H\beta)}<2000\ \mathrm{km\ s^{-1}}$)
to date (Oknyansky et al. 2018; Xu et al. 2024, and references therein; Yang et al. 2025). Yang et al. (2025) claimed that the probable fraction of CL-NLS1s is not larger than 7\% in their sample.
Owing to their small $M_\mathrm{BH}$ and high $L/L_{\mathrm{Edd}}$, NLS1s are believed to be
``young'' AGNs with not only rapid SMBH growth caused 
by accretion (e.g., Boroson 2002; Xu et al. 2012), but also strong circumnuclear star formation (e.g., Mathur 2000; Sani et al. 2010; Scharwachter et al. 2017; Orban de Xivry et al. 2011; Mathur et al. 2012). 

Does the intermediate-old stellar population suggested for CL-AGNs by previous studies
in fact result from a sample selection effect? Based on the
large NLS1 sample recently compiled by Paliya et al. (2024), in this study
we perform a systematic search for CL-NLS1s to verify the evolutionary 
role of CL-AGNs by extending our study to the high $L/L_{\mathrm{Edd}}$ and young 
stellar population ends.
The new 
NLS1 sample contains 22,656 NLS1s extracted from the spectra in 
SDSS Data release 17. Involving the 
multi-epoch spectroscopy in the SDSS-V DR19 data release enables us to identify 
CL-NLS1s, and to investigate the properties of the associated circumnuclear 
stellar population in their low-state (``turn-off'') spectra. This approach represents a
unique way to study the SMBH environment in NLS1s, unaffected by the bright high-state continuum and BLR emission 
that hampers such studies in the turn-on state.

The paper is organized as follows. 
The search for CL-NLS1s is presented in Section 2. Section 3 gives the spectral 
modeling. The analysis and results inferred from the modeling are presented in Section 4 
Our conclusions and the implications for the evolutionary role of CL-AGNs are discussed in Section 5, along with an extension to 
the potential value of examining the narrow-line region (NLR) of type-I AGNs by their
off-center spectra. 
A $\Lambda$CDM cosmological model with parameters H$_0=70\,\mathrm{km\,s^{-1}\,Mpc^{-1}}$, $\Omega_{\mathrm{m}}=0.3$, and
$\Omega_\Lambda=0.7$ is adopted throughout the paper.



\section{Searching for CL-NLS1 Galaxies} \label{sec:style}

\subsection{SDSS-V DR19}

The fifth phase of the Sloan Digital Sky Survey (SDSS-V)
is  the first all-sky, multi-epoch, optical-to-infrared spectroscopic survey
(Kollmeier et al. 2025). The survey is carried out through multi-object spectroscopy
with telescopes in both hemispheres, and consists of three primary survey programs: Milky Way Mapper, Black Hole Mapper (BHM), and Local Volume Mapper. The main purpose of the BHM is to study AGNs over long time intervals to measure black hole masses and discover CL quasars, and to optically characterize {\it eROSITA} X-ray sources. 
Based on the SDSS DR16 quasar catalog (Lyke et al. 2020), 
the All-Quasar Multi-Epoch Spectroscopy program 
will provide multi-epoch optical ($\lambda = 360$-–1000~nm)
spectra of $\sim 20,000$ QSOs with $i<19.1$~mag at a spectral resolution of 
$R \equiv \lambda/\Delta\lambda \approx 2000$ (Smee et al. 2013) that is comparable to that of 
Baryon Oscillation Spectroscopic Survey. 
The number of epochs ranges from a few to a dozen on baselines from months to 
a decade. In order
to cover the whole sky, dual-hemisphere observations are performed in a parallel way with two 2.5~m telescopes. One is the 2.5~m Sloan Foundation Telescope (Gunn et al. 2006) at Apache Point Observatory (APO) in New Mexico, and the other is the 2.5~m 
du Pont telescope of the Carnegie Observatories at Las Campanas Observatory (LCO) in Chile (Bowen \& Vaughan 1973). The nineteenth data release (DR19) of SDSS became 
recently available for all three mappers (SDSS Collaboration et al. 2025).

\subsection{Cross-Match and Identification}

We start from the catalog of NLS1s recently provided by Paliya et al. (2024). 
As a first step, the redshifts of the NLS1s are required to be $< 0.45$ 
to ensure the H$\alpha$ emission line can be covered by the SDSS
spectroscopic wavelength range in the observer frame. The remaining objects are 
then cross-matched with the SDSS DR19 BHM catalog (dataset v6\_1\_3) by their celestial coordinates within a cross-matching radius of 3\arcsec; 
this returns 3705 matches corresponding 884 individual objects. 

For each of the 3705 matches, a differential spectrum is used to search for new 
CL-NLS1, and is created from the corresponding SDSS DR19 and DR16 spectra as follows.
First, both SDSS DR19 and DR16 spectra for each match are transformed to 
the rest frame according to the redshift reported in the SDSS DR16 catalog, 
along with a correction of Galactic extinction by adopting an $R_V = 3.1$
extinction law of the Milky Way (Cardelli et al. 1989). 
The Galactic reddening map of Schlegel, Finkbeiner, \& Davis (1998) is used for 
extracting the corresponding extinction values.

For both spectra in each match, the local continuum of the [\ion{O}{3}]$\lambda5007$ narrow emission line
is modeled by a linear function based on the line-free wavelength range 
around the line. After removing the modeled local continuum from the observed spectrum. 
the [\ion{O}{3}]$\lambda5007$ line flux is measured by a direct integration in 
the wavelength range 4995--5019\,\AA. By using the SDSS DR16 spectrum as 
a reference, a differential spectrum is  
obtained after scaling the flux level of the SDSS DR19 spectrum by the measured 
[\ion{O}{3}] line flux ratio $f_{\mathrm{DR19}}/f_{\mathrm{DR16}}$. 

After removing the underlying ``continuum'' from the differential spectrum by 
a low-order polynomial function, a CL-NLS1 candidate is identified by requiring 
$f(\mathrm{H\beta}) > 3\sigma_{\mathrm{c}}$, where $f(\mathrm{H\beta})$ is 
the mean flux of the H$\beta$ emission line measured in the range 4850--4872\,\AA\ and 
$\sigma_{\mathrm{c}}$ is the fluctuation of the ``continuum''-removed differential spectrum 
over 5100--5300\,\AA. The spectra of the identified candidates are inspected 
individually by eye,  enabling us to identify  three CL-NLS1s from the 
884 individual objects. The three CL-NLS1s are SDSS~J080643.25+153815.2, SDSS~J084637.64+000024.4\footnote{The object has also been identified as a CL-AGN by Zeltyn et al. (2024).}, \hfill and\\ 
SDSS~J142135.90+523138.9\footnote{The H$\beta$ line width is reported to be 
$\mathrm{FWHM(H\beta)}=1830\pm40$, $1720\pm110$ and $1520\pm40\ \mathrm{km\ s^{-1}}$ for 
the three objects by Paliya et al. (2024).}. In addition, although most of the candidates are caused by a 
strong line variation, $\sim30$ candidates are found to have a large value of 
$\Delta f(\mathrm{[O~III]]})/f_{\mathrm{DR16}} > 0.5$, where $\Delta f(\mathrm{[O~III]} = 
|f_{\mathrm{DR19}}-f_{\mathrm{DR16}}|$, 
which potentially resulted from the ``fiber drop'' effect 
(e.g., Guo et al. 2025) and will be addressed in Section 5.4 below.

\begin{figure*}[htp!]
\plotone{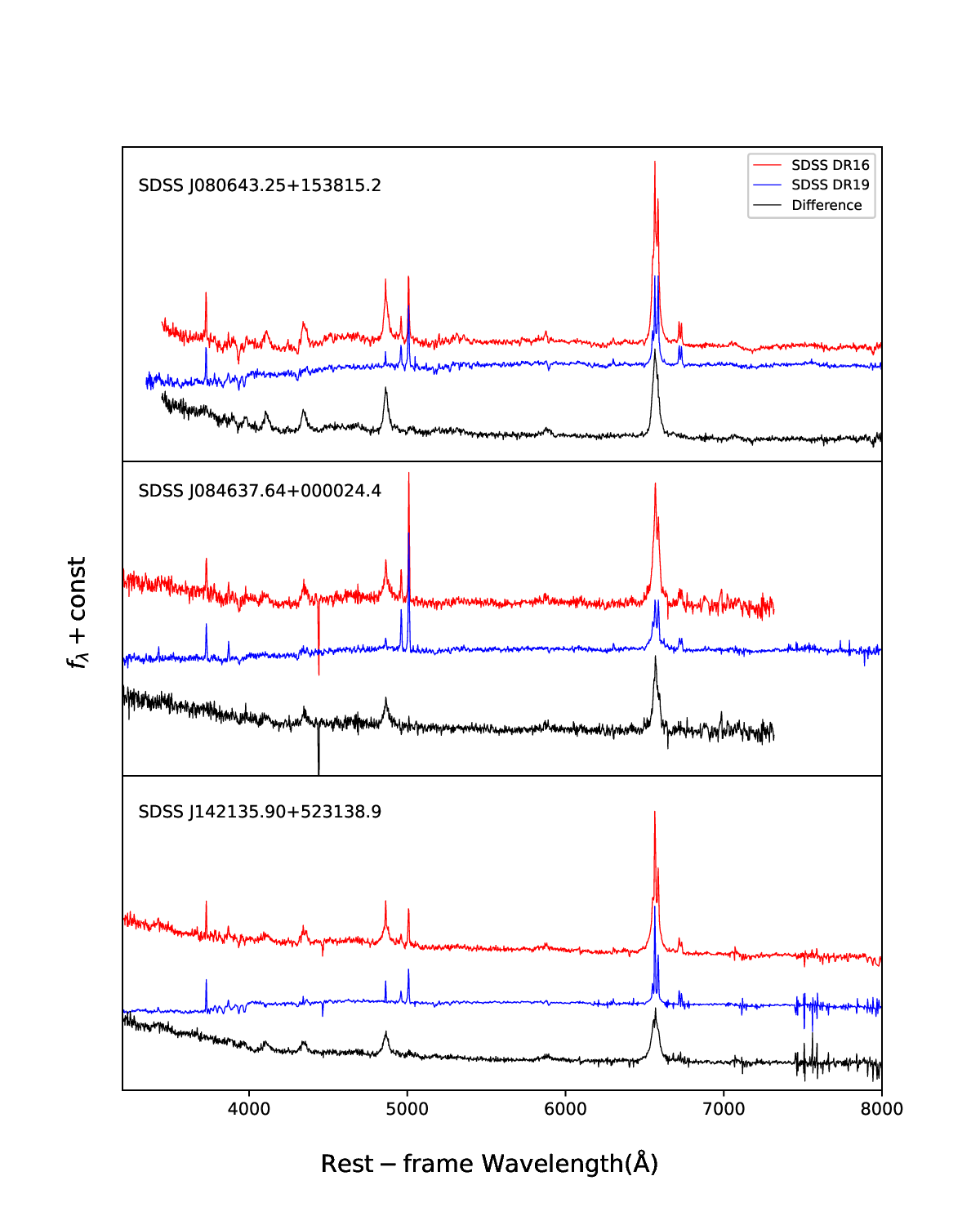}
\caption{Identification of the three identified CL-NLS1s by comparing their SDSS DR16 and SDSS DR19 spectra.  
In each panel, the SDSS DR16, SDSS DR19 (in which the multi-epoch SDSS DR19 spectra are combined) 
and the corresponding differential spectra are shown from 
top to bottom by the red, blue and black lines, respectively, 
The differential spectrum is generated 
by using the SDSS DR16 spectrum as a reference. The SDSS DR16 and SDSS DR19 are 
scaled to a common level according to their [\ion{O}{3}] $\lambda$5007 fluxes before the subtraction.
The spectra are binned by a boxcar of 3\,\AA\ and shifted vertically by an arbitrary amount for clarity. 
\label{fig:diff}}
\end{figure*}

The identification of the three CL-NLS1s is illustrated in Figure \ref{fig:diff} by 
comparing the corresponding SDSS DR19, SDSS DR17 and differential spectra in the rest frame.
In two cases where there are multi-epoch spectra in the SDSS DR19 dataset, all these
spectra are combined for subsequent spectral analysis, after confirming they are 
at the ``turn-off'' state by an examination by eyes. The number and the corresponding
time span of the combined SDSS DR19 spectra are listed in Table \ref{tab:properties}.    
As shown in the figure, the CL phenomenon can be clearly identified in the three objects 
owing to the disappearance of the H$\beta$ broad-line emission in their SDSS DR19 spectra. That means
the SDSS DR19 spectra were taken in the ``turn-off'' state. All three differential spectra are
characterized by not only blue featureless continuum, but also strong broad Balmer line emission,
which allows us to assess broad Balmer line flux variation by a direct integration of the three differential spectra by the 
IRAF/SPLOT task\footnote{IRAF is distributed by NOAO, which is 
operated by AURA, Inc., under cooperative agreement with the U.S. National Science Foundation (NSF).}, 
The measurements, along with the timescale of ``turn-off'' determined from the epochs of the spectra,
are presented in Table 1\footnote{The statistical uncertainty of the line flux reported in
the table is determined by the method given by Perez-Montero \& Diaz (2013): 
$\sigma_\mathrm{l}=\sigma_\mathrm{c}N[1+EW/(N\Delta)]$, where $\sigma_\mathrm{c}$ and 
$EW$ are the standard deviation of the continuum in a box near the line and equivalent
width of the line, respectively. $\Delta$ denotes the wavelength dispersion in units of 
\AA\,pixel$^{-1}$. },  

\begin{table*}
        \centering
        \caption{Measurements of the Three Differential Spectra}
        \footnotesize
        \label{tab:example_table}
        \begin{tabular}{ccccccccccc} 
        \hline
        \hline
        SDSS ID & $z$ & $\Delta t$ & Number & MJD span & $\mathrm{H\alpha}$ &  $\mathrm{H\beta}$ & $\mathrm{H\gamma}$  & $\mathrm{H\delta}$  & $\mathrm{H\epsilon}$ &  He~I $\lambda5876$ \\
        & & (yr) & & & \multicolumn{6}{c}{($10^{-15}\ \mathrm{erg\ s^{-1}\ cm^{-2}}$)} \\
        (1) & (2) & (3) & (4) & (5) & (6) & (7) & (8) & (9) &  (10) & (11)\\
        \hline 
        080643.25+153815.2 & 0.105 & 13.2 & 3 & (59874, 59879) & $28.2\pm0.2$ & $10.5\pm0.1$ &  $5.8\pm0.1$  & $4.9\pm0.1$ & $1.7\pm0.1$  &  $1.8\pm0.1$\\
        084637.64+000024.4 & 0.257 & 16.9 & 1 & 59227 & $7.2\pm0.1$  & $2.0\pm0.6$  &  $0.8\pm0.1$  & & & \\        
        142135.90+523138.9 & 0.249 & 4.0 & 44 & (59636, 60118) & $5.9\pm0.1$ & $1.8\pm0.1$  &  $0.9\pm0.1$ & $0.7\pm0.1$ &                $0.3\pm0.1$ & \\
        \hline
        \end{tabular}
        \tablecomments{Column (1): SDSS identification; Column (2): redshift; Column (3): timescale of ``turn-off'' in units of years. Column (4): number of the combined SDSS DR19 spectra; Column (5):  time span in MJD of the combined SDSS DR19 spectra. Columns (6-11): 
        line fluxes measured in the differential spectrum.}
        \label{tab:properties}
\end{table*}

\section{Spectral Modeling}

There are two motivations in this study. The first  is 
the identification of CL-NLS1s and measurement of their properties, 
since these rare events have important implications for accretion-disk 
physics (e.g., Xu et al. 2024). The second is to explore the 
evolutionary role of CL-AGNs by extending our previous study to CL-NLS1s,
based upon the
properties of the associated circumnuclear stellar population. 
With these motivations, the spectra are analyzed as follows.

\subsection{The ``Turn-off'' Spectra}

\subsubsection{Continuum}

As shown in Figure \ref{fig:diff}, since the continuum of the
SDSS DR19 ``turn-off'' spectra of the three CL-NLS1s are dominantly (or mainly) contributed 
by their host galaxies,  we model the continuum of the ``turn-off'' spectra by a  
linear combination of the starlight component and an underlying AGN contribution, 
along with an intrinsic extinction. 

The AGN's contribution is modeled by a scaled power law, $f_\lambda\propto\lambda^{-\alpha}$,
with a fixed index of $\alpha =1.7$ determined by the mean spectrum of quasars (e.g., Vanden Berk et al. 2001, and references therein).
By following Wang et al. (2023), the starlight component is described by a linear combination of the first
seven eigenspectra that are built through the principal component analysis method (e.g., Francis et al. 1992; Hao
et al. 2005; Wang \& Wei 2008) from the standard single stellar population spectral library developed by 
Bruzual \& Charlot (2003). The extinction is described by a Galactic extinction curve with $R_V = 3.1$.

The continuum in each spectrum is fitted iteratively by a $\chi^2$ minimization, 
which covers the whole wavelength range, except the regions with known strong emission lines.
These regions include the broad low-order Balmer lines, [\ion{S}{2}]$\lambda\lambda$6716, 6731,
[\ion{N}{2}]$\lambda\lambda$6548, 6583, [\ion{O}{1}]$\lambda$6300, [\ion{O}{3}]$\lambda\lambda$4959, 5007, 
[\ion{O}{2}]$\lambda\lambda$3726, 3729, [\ion{Ne}{3}]$\lambda$3869, and
[\ion{Ne}{5}]$\lambda$3426. The starlight velocity dispersion is not fixed in the fitting.
The left panels of Figure \ref{fig:modeling_off} show the modeled continuum and emission-line isolated spectra
for the three CL-NLS1s.

\subsubsection{Emission-Line Profile}

\begin{figure*}[htp!]
\plotone{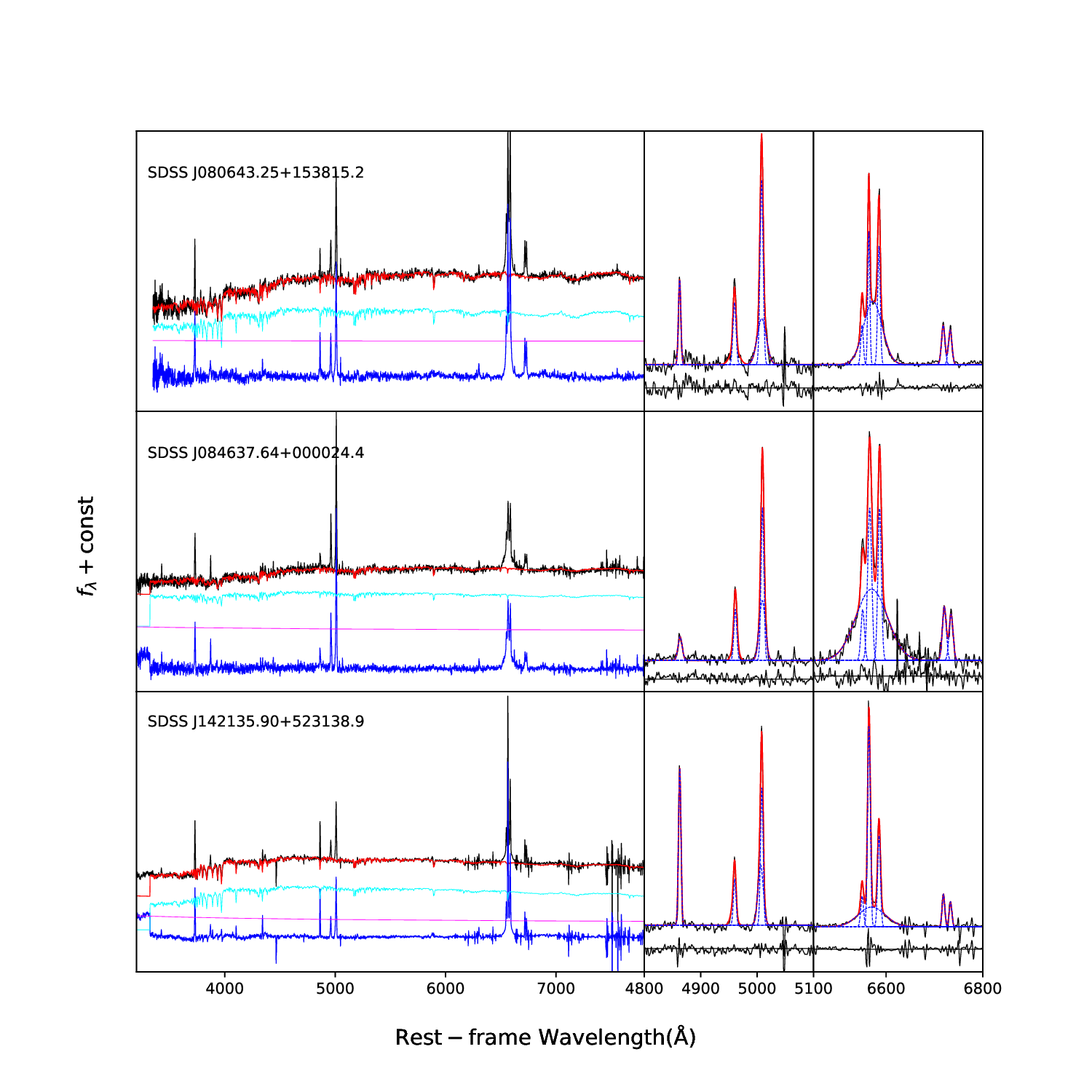}
\caption{Spectral decomposition and emission-line profile fitting. \it Left panels: \rm The continuum 
modeling process. In each panel, the observed rest-frame spectrum (the heavy black curve) is fitted 
by a modeled being consist of a starlight component (the cyan curve) and an AGN's powerlaw with a free slope (the magenta curve). The blue curve at the bottom is the resulted emission-line spectrum after 
a subtraction of the best-fit continuum (the red curve).  
All spectra are shifted vertically by an arbitrary amount for clarity.
{\it Middle panels:}  Line-profile modeling for the the H$\beta$ region in the emission-line isolated 
spectra (see the left panels). The modeling is based on a a linear combination of a set of
Gaussian functions. The observed and
modeled line profiles are shown by black and red solid lines, respectively. Individual Gaussian function resulted from the fitting is shown as blue dashed lines. 
The residuals between the observed and modeled profiles are presented in 
the subpanel below.  
{\it Right panels:}  Identical in layout to the middle panels, but for the H$\alpha$ region.
\label{fig:modeling_off}}
\end{figure*}

The IRAF/SPECFIT task (Kriss 1994) is employed   
to fit the emission-line profiles in both the H$\alpha$ and H$\beta$ by 
a linear combination of Gaussian functions to the emission-line isolated spectra, in which the theoretical values of 1:3 are predefined for the 
[\ion{O}{3}] $\lambda\lambda$4959, 5007 and [\ion{N}{2}]$ \lambda\lambda$6548, 6583 doublets ratios.
For the [\ion{O}{3}]$ \lambda\lambda$4959, 5007 line profiles, both a narrow component and a 
blueshifted broad component are necessary for providing a properly modeling 
(e.g., Boroson 2005; Wang et al. 2011, 2018; Zhang et al. 2013; Harrison et al. 2014; Woo et al. 2017).
Similarly, the H$\alpha$ line profiles are best reproduced with both a narrow and broad components. 
On the contrary, only a narrow component can adequately reproduce the H$\beta$ line profile properly 
in all three CL-NLS1s.
The middle and right panels of Figure \ref{fig:modeling_off} displays the best line-profile modeling 
for the H$\beta$ and H$\alpha$ regions, respectively. 

\subsection{The ``Turn-on'' Spectra}

\begin{figure*}[htp!]
\plotone{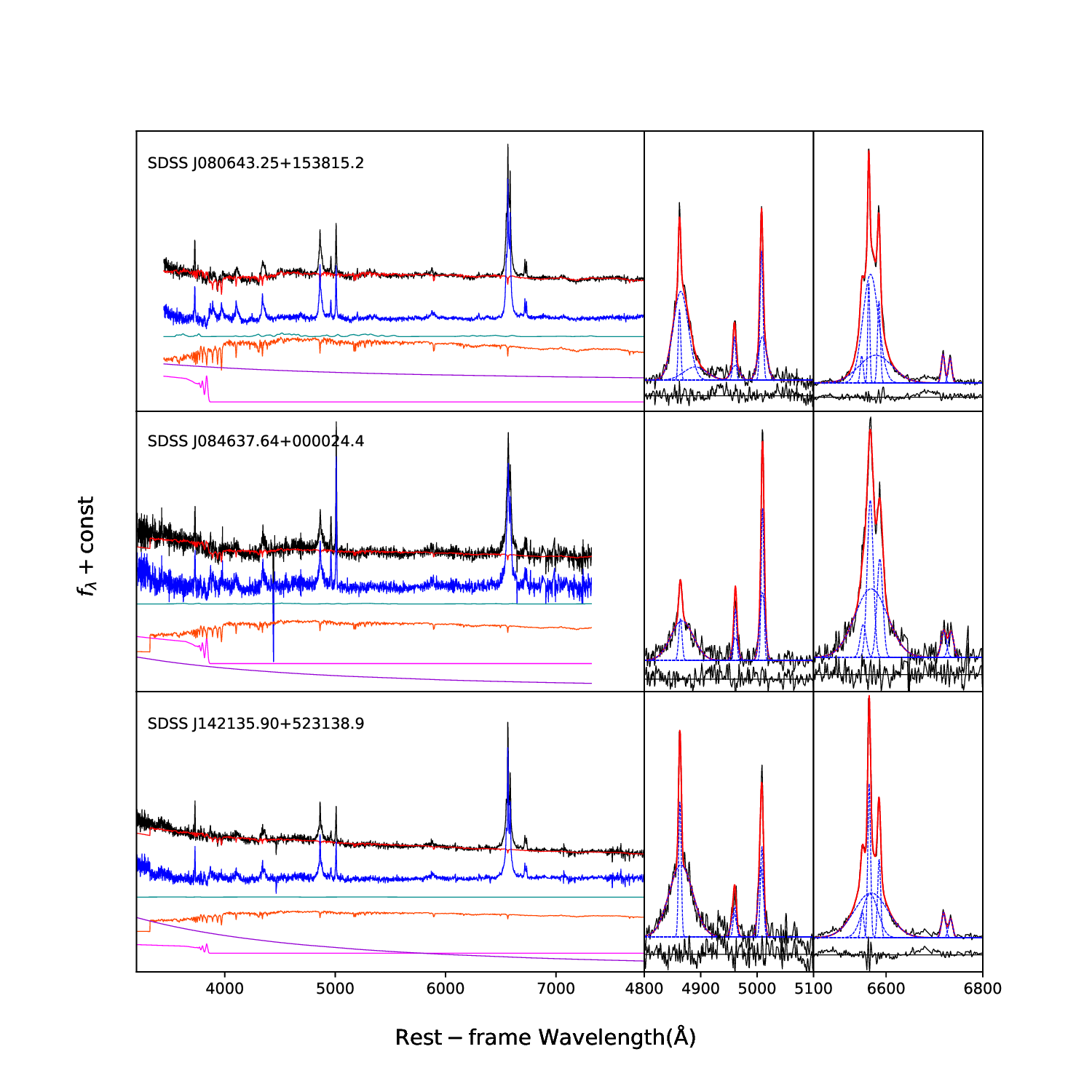}
\caption{The same as Figure \ref{fig:modeling_off}, but for the ``turn-on'' spectra. As shown in the left panels, in addition to the AGN's continuum and the underlying starlight component, the Balmer continuum and high-order Balmer lines (magenta curve), along with the \ion{Fe}{2} complex (the green curve), are 
required to model the continuum properly in each spectrum. 
\label{fig:modeling_on}}
\end{figure*}

The continuum of each of the ``turn-on'' spectra is analyzed by a composite spectral model, which is 
parameterized as a linear combination of the following components: (1) an AGN power-law continuum with a 
free slope index, (2) a template of
both high-order Balmer emission lines and a Balmer continuum from the BLR, (3)
a template of the optical \ion{Fe}{2} complex, (4) a scaled host-galaxy component obtained
from the corresponding ``turn-off'' spectrum, and (5) an intrinsic extinction
due to the host galaxy following the Galactic extinction curve with $R_V = 3.1$.
Similar to for the ``turn-off'' spectra, the fitting is performed via $\chi^2$ minimization 
over the entire spectral wavelength range, except the regions with known strong emission lines.
The continuum modeling and removement is presented in the left panels of Figure \ref{fig:modeling_on}.

The empirical optical \ion{Fe}{2} templates given in Veron-Cetty et al. (2004),
including both broad and narrow components of the \ion{Fe}{2} emission, 
is adopted to model the optical \ion{Fe}{2} complex. 
The line widths of both templates are fixed in advance to be those of the broad and narrow 
components of H$\beta$
by convolving the templates with a Gaussian profile whose width is determined by our 
line-profile modeling.

By following Dietrich et al. (2002; see also Malkan \& Sargent 1982), we model the 
Balmer continuum $f^{\mathrm{BC}}_\lambda$ by 
the emission from a partially optically thick cloud:

\begin{equation}
  f^{\mathrm{BC}}_\lambda=f^{\mathrm{BE}}_\lambda B_\lambda(T_{\rm e})(1-e^{-\tau_\lambda}), \  \lambda<\lambda_{\mathrm{BE}}\, ,
\end{equation}
where $f^{\mathrm{BE}}_\lambda$ is the continuum flux at the Balmer edge ($\lambda_{\mathrm{BE}} = 3646$\,\AA), and $B_\lambda(T_{\rm e})$ the Planck function. 
$\tau_\lambda$ is the optical depth at wavelength $\lambda$, 
and related to the one at $\lambda_{\mathrm{BE}}$ by $\tau_\lambda=\tau_{\mathrm{BE}}(\lambda/\lambda_{\mathrm{BE}})^3$. 
The typical values of $\tau_{\mathrm{BE}}=0.5$ and $T_{\rm e} = 1.0\times10^4$~K are adopted in our continuum modeling.

We model the high-order Balmer lines (i.e., H7--H50) by the Case B recombination model that is
parameterized by
$T_{\rm e} = 1.5\times10^4$~K and an electron density of $n_{\rm e} = 10^{7-8}\ \mathrm{cm^{-3}}$ 
(Storey \& Hummer 1995). Similar as for the \ion{Fe}{2} complex,
the widths of these high-order Balmer lines are, again, fixed in advance according to the line-profile modeling
of the H$\beta$ broad emission.

After subtracting the fitted continuum from the observed ``turn-on'' spectra, 
we model the line profiles in both the H$\beta$ and H$\alpha$ regions by following the 
method described in Section 3.1.2. The fitting of the line profiles is 
shown in the middle and right panels of Figure \ref{fig:modeling_on}.

\vspace{0.5cm}

The spectral modeling results are summarized in Table \ref{tab:analysis_results}.  
When the broad Balmer emission has to be reproduced by two Gaussian functions,
we at first create a residual line profile by subtracting the modeled narrow component from the 
observed one, and then measure the FWHM from the residual profile.

\begin{longrotatetable}
\begin{deluxetable*}{ccccccccccccc}
\tablecaption{Results of Spectral Analysis of the Three CL-NLS1s\label{tab:analysis_results}}
\tabletypesize{\footnotesize}
\tablehead{
\colhead{SDSS ID} &
\colhead{$z$}  &
\colhead{$F(\mathrm{[OIII}\lambda5007)$} &
\colhead{$F(\mathrm{H\beta_b})$} &
\colhead{$F(\mathrm{H\alpha_{\mathrm{b}}})$} &  
\colhead{$\mathrm{FWHM(H\beta_{b})}$}  & 
\colhead{$\mathrm{FWHM(H\alpha_{b})}$}  &
\colhead{R4570} &
\colhead{$\log(M_{\mathrm{BH}}/{\rm M}_\odot)$} &
\colhead{$L/L_{\mathrm{Edd}}$} & 
\colhead{$D_{\mathrm{n}}(4000)$} &
\colhead{H$\delta_{\mathrm{A}}$} & 
\colhead{Status}
\\
\colhead{} & \colhead{}  &
\multicolumn{3}{c}{$\mathrm{(10^{-15}\ erg\ s^{-1}\ cm^{-2})}$} &
\multicolumn{2}{c}{$\mathrm{(km\ s^{-1})}$} & & &   & & (\AA) & \\
}
\colnumbers
\startdata
0806+1538 & 0.105 & $5.3\pm0.3$ & $11.8\pm0.7$ & $43.0\pm3.4$ & $2080\pm40$ & $1840\pm60$ & $0.47\pm0.03$ & $7.06\pm0.03$ & $0.16\pm0.02$ & & &  on\\
& & $4.7\pm0.3$ & & $11.8\pm0.2$ & &  $2070\pm20$ & &  & $0.05\pm0.01$ & $1.34\pm0.08$ & $0.83\pm0.01$ & off\\
0846+0000 & 0.257 & $2.8\pm0.6$ & $3.2\pm0.1$ & $8.6\pm0.3$ & $2920\pm140$ &  $3660\pm110$  & $0.12\pm0.01$ & $8.08\pm0.03$ & $0.10\pm0.01$ & & &  on\\
& & $2.9\pm0.3$ & & $3.8\pm0.1$ & & $3530\pm140$ &  &  & $0.01\pm0.002$ & $1.43\pm0.08$ & $-2.20\pm0.42$ & off\\
1421+5231 & 0.249 & $0.8\pm0.1$ & $2.2\pm0.1$ & $8.3\pm0.6$ & $3070\pm100$ & $2480\pm130$ & $0.04\pm0.01$ & $7.42\pm0.05$ & $0.11\pm0.01$ & & &  on\\ 
& & $0.6\pm0.1$ & & $1.5\pm0.1$ & & $2960\pm130$ &  & & $0.04\pm0.01$ & $1.24\pm0.04$ & $2.20\pm0.36$ & off\\
\enddata
\tablecomments{Columns (1) \& (2): SDSS identification and the corresponding redshift;
Columns (3-5): the measured line fluxes of [\ion{O}{3}]$\lambda5007$, $\mathrm{H\beta}$ 
broad component and $\mathrm{H\alpha}$ broad component; Columns (6) and (7): line widths of 
the broad components of $\mathrm{H\beta}$ and  $\mathrm{H\alpha}$; Column (8): 
\ion{Fe}{2} strength $\mathrm{R4570 = FeII/H\beta}$; Column (9): blackhole mass estimated 
from the ``turn-on'' spectrum by Eq. (3);  
Column (10): Eddington ratios at both ``turn-on'' and
``turn-off'' states; Columns (11) \& (12): Lick indices measured in the ``turn-off'' spectrum; 
Column (13): CL state assessed by the spectrum. All the uncertainties correspond to 1$\sigma$ 
significance level.} 
\end{deluxetable*}
\end{longrotatetable}

\section{Analysis and Results}
\subsection{Eddington Ratio}

Based on the modeled broad H$\alpha$ emission, we calculate the
Eddington ratio $L_{\mathrm{bol}}/L_{\mathrm{Edd}}$ ($L_{\mathrm{Edd}}=1.5\times10^{38}\,(M_{\mathrm{BH}}/{\it M}_\odot)\,\mathrm{erg\,s^{-1}}$ is the Eddington luminosity; see Eq. 3.16 of Netzer 2013) at their ``turn-off'' and ``'turn-on'' states. The bolometric luminosity 
$L_{\mathrm{bol}}$ is inferred by using the widely used bolometric correction of $L_{\mathrm{bol}}=9\lambda L_{\lambda}(5100\,{\rm \AA})$  (e.g., Kaspi et al. 2000), in which 

\begin{equation}
 \lambda L_\lambda(5100~\mathrm{\AA}) = 2.4\times10^{43}\bigg(\frac{L_{\mathrm{H\alpha}}}{10^{42}\ \mathrm{erg\ s^{-1}}}\bigg)^{0.86}\ \mathrm{erg\ s^{-1}}\, ,
\end{equation}
according to the $L_{5100~{\rm \AA}} –L_{\mathrm{H\alpha}}$ relationship calibrated by Greene \& Ho 
2005). The modeled broad H$\alpha$ line enables us to calculate the black hole mass $M_{\mathrm{BH}}$ 
and the uncertainty for each of the three objects by following the calibration reported by Greene \& Ho (2007),
\begin{equation}
  M_{\mathrm{BH}} = 3.0\times10^6\bigg(\frac{L_{\mathrm{H\alpha}}}{10^{42}\ \mathrm{erg\ s^{-1}}}\bigg)^{0.45}
  \bigg(\frac{\mathrm{FWHM(H\alpha)}}{1000\ \mathrm{km\ s^{-1}}}\bigg)^2M_\odot \, .
\end{equation}
In both equations, $L_{\mathrm{H\alpha}}$ is the H$\alpha$ line luminosity corrected for
the intrinsic extinction that is inferred from the Balmer decrement of $\mathrm{H\alpha/H\beta}$. 
The standard Case B recombination, along with a Galactic extinction curve with $R_V=3.1$, is 
involved in the extinction correction.

Column (10) in Table 2 compares the values of $L_{\mathrm{bol}}/L_{\mathrm{Edd}}$, along with their uncertainties, \rm estimated at 
the ``turn-on'' and ``turn-off'' states, where the $M_{\mathrm{BH}}$ evaluated from the 
``turn-on'' spectra is used to calculate the $L_{\mathrm{bol}}/L_{\mathrm{Edd}}$ at the 
``turn-off'' state. 
One can see from the comparison that the $L_{\mathrm{bol}}/L_{\mathrm{Edd}}$ at the 
``turn-on'' state is higher than that at the ``turn-off'' state by a factor of 3--10 times,  
much larger than the typical uncertainty of $\sim65\%$ (or 0.28~dex) mainly caused by the calibration scatter.

\subsection{Stellar Population of the Hosts}

Despite their known sensitivity to metallicity in very old stellar populations,
both Lick indices, [$D_{\rm n}(4000)$] and $\mathrm{H\delta_A}$, have been severed as 
reliable age indicators until a few Gyr after a starburst in literature (e.g., Kauffmann et al. 2003; Heckman et al. 2004; Kewley et al. 2006; Kauffmann \& Heckman 2009; Wild et al. 2010; Wang \& Wei 2008, 2010; Wang et al. 2013; Wang 2015).

The index $D_{\rm n}(4000)$ (the 4000~\AA\ break) is defined as (Balogh et al. 1999; Bruzual 1983)
\begin{equation}
 D_{\rm n}(4000)=\frac{\int_{4000}^{4100}f_\lambda d\lambda}{\int_{3850}^{3950}f_\lambda d\lambda}\, .
\end{equation}
and the index H$\delta_{\mathrm A}$ (the equivalent width (EW) of the H$\delta$ absorption due to A-type stars) as (Worthey \& Ottaviani 1997)
\begin{equation}
   \mathrm{H\delta_A}=(4122.25–4083.50)\bigg(1-\frac{F_I}{F_c}\bigg)\ \mathrm{\AA}\, ,
\end{equation}
where $F_I$ is the flux within the feature bandpass 4083.50--4122.25~\AA,
and $F_c$ is the flux of the pseudocontinuum evaluated in the two adjacent regions,
blue (4041.60--4079.75~\AA) and red (4128.50--4161.00~\AA).

Both indices are measured from the starlight components decomposed from the ``turn-off'' 
spectra, and tabulated in Table 2. The listed uncertainties are obtained by 
a boostrap estimation, and is comparable to
the typical uncertainties of $\Delta D_{\rm n}(4000)\sim0.04$ and $\Delta \mathrm{H\delta_A}\sim0.4$\AA\
assessed from the duplicated spectra of Seyfert 2 galaxies in Wang (2015).

\subsection{Statistics}

\begin{figure*}[ht!]
\begin{tabular}{cc}
   \begin{minipage}{0.50\linewidth}
      \centering
      \plotone{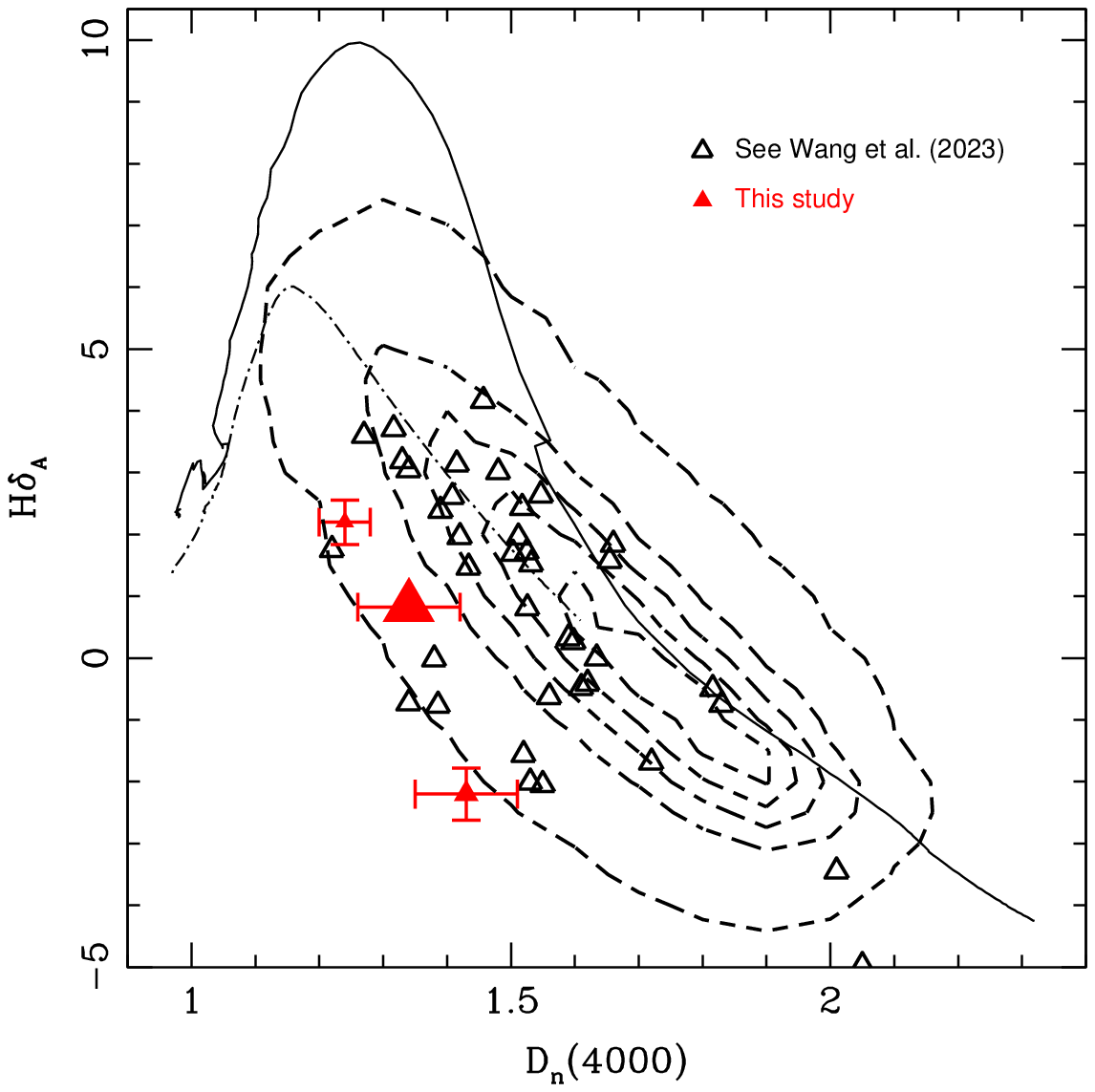}
   \end{minipage} &
   \begin{minipage}{0.50\linewidth}
      \centering
      \plotone{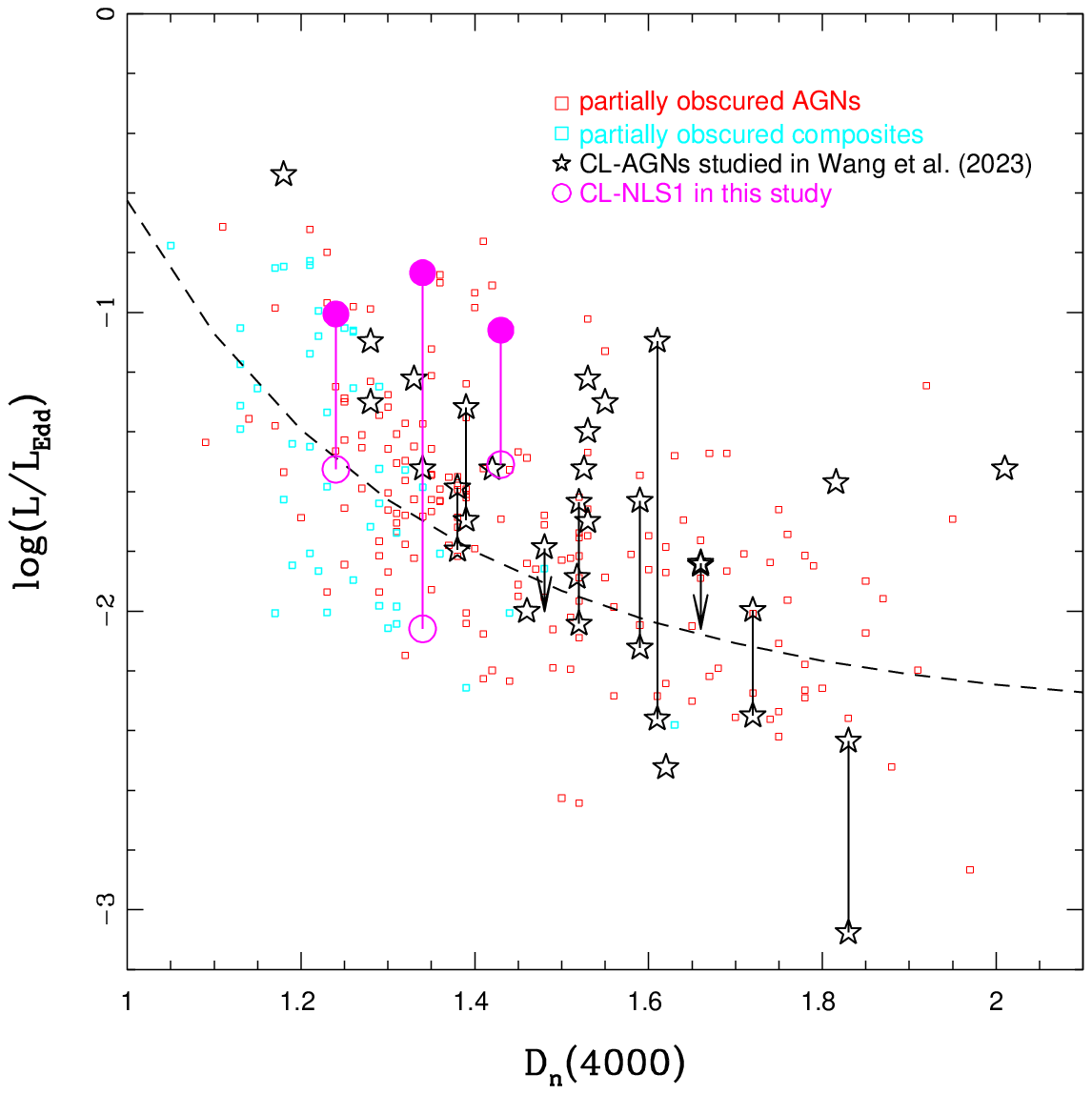}
   \end{minipage}  \\
   \end{tabular}
\caption{\it Left panel: \rm $D_{\rm n}(4000)$ plotted as a function of $\mathrm{H}\delta_{\mathrm{A}}$ for local CL-AGNs, including both the 
three CL-NLS1s (the red solid triangles) reported in this study and the CL-AGNs (the black open triangles) compiled by 
Wang et al. (2023). For the three CL-NLS1s, the size of the solid triangles depends on the value of  $\mathrm{R4570 = Fe~II/H\beta}$ reported by 
Paliya et al. (2024).  As a comparison, the dashed black heavy lines correspond to the density contours  of distribution of
 $\sim 80,000$ Seyfert 2 galaxies provided in the 
MPA/JHU value-added catalog (e.g., Kauffmann et al. 2003; Heckman \& Kauffmann 2006). 
The stellar population evolution locus are over plotted in the plot by the solid line for  the single-stellar-population model
with solar metallicity, and by the dot-dashed line for the exponentially decreasing star-formation rate $\psi(t)\propto e^{-(t/\mathrm{4~Gyr})}$.
\it Right panel: \rm Distribution on the $L_{\mathrm{bol}}/L_{\mathrm{Edd}}-D_{\rm n}(4000)$ diagram.  
The three 
CL-NLS1s are  shown by the magenta circles, and the local CL-AGNs compiled by Wang et al. (2023)  by the black open stars. 
The ``turn-off'' and ``turn-on'' states are connected by a vertical line for individual CL-AGNs. 
The local ``partially obscured'' AGNs studied in Wang (2015) are shown by the small squares for a comparison 
(red: Seyfert galaxies, cyan: composite galaxies). 
The best-fit nonlinear relationship to these ``partially obscured'' AGNs is presented by the dashed line.
\label{fig:stellar}}
\end{figure*}

The left panel in Figure \ref{fig:stellar} shows the H$\delta_{\mathrm{A}}$ versus $D_{\rm n}(4000)$ plot, which compares the 
three CL-NLS1 sample and the $\sim60$ local CL-AGN sample previously studied by Wang et al. (2023).
Two facts can be learned from the figure. On the one hand,  although
the three CL-NLS1s are biased toward  ``young'' stellar populations\footnote{A value of $D_{\rm n}(4000) = 1.4$-–1.6 is
usually adopted for separating young and old stellar populations (e.g., Kauffmann et al. 2003).} with relatively
small $D_{\rm n}(4000)$ ($1.20<D_{\rm n}(4000) <1.45$), the three CL-NLS1s are consistent with the
general  $\mathrm{H\delta_A}-D_{\rm n}(4000)$ sequence of the CL-AGNs reported in Wang et al. (2023).
By combining the three CL-NLS1s with the CL-AGNs studied in Wang et al. (2023), 
the average value of $D_{\rm n}(4000)$ is determined to be 
$\langle D_{\rm n}(4000)\rangle = 1.51\pm0.18$ for the combined sample, which is comparable 
to the values of $\langle D_{\rm n}(4000)\rangle = 1.52\pm0.18$ for the CL-AGNs studied in
Wang et al. (2023). Comparing the $D_{\rm n}(4000)$ 
of all CL-AGNs shown in the figure and of the Seyfert 2 galaxies in the MPA/JHU catalog 
(e.g., Kauffmann et al. 2003; Heckman \& Kauffmann 2006), the one-dimensional two-sided
Kolmogorov-Smirnov (KS) test yields that the two distributions are different with a 
$p$-value of $3.2\times10^{-7}$ at a maximum distance of 0.34. 
The similar KS-test in two dimensions (Peacock 1983; Fasano \& Franceschini 1987), 
 $D_{\rm n}(4000)$ and H$\delta_{\mathrm{A}}$, yields that the 
stellar populations of the host galaxies of the CL-AGNs differ from those
of the MPA/JHU catalog at a maximum distance of 0.37 with a $p$-value of $1.2\times10^{-6}$.  
On the other hand, there is a tendency that CL-AGNs are separated into two groups with
different H$\delta_{\mathrm{A}}$ at 
a fixed $D_{\rm n}(4000)$, which needs to be confirmed by larger samples in the future.

The right panel in Figure \ref{fig:stellar}  marks the three CL-NLS1s in the $D_{\rm n}(4000)$-–$L_{\mathrm{bol}}/L_{\mathrm{Edd}}$ diagram, 
A relationship between $L_{\mathrm{bol}}/L_{\mathrm{Edd}}$ and $D_{\rm n}(4000)$ has been identified in 
a series of previous studies of local AGNs, in which  $L_{\mathrm{bol}}/L_{\mathrm{Edd}}$ decreases as the 
young stellar population continuously ages
(e.g., Kewley et al. 2006; Wang et al. 2006, 2013; Wang \& Wei 2008, 2010). 
A direct interpretation of this relationship is a coevolution between SMBH growth and host galaxy wherein the SMBH resides.  
One can see from the plot that 
the three CL-NLS1s occupy the  upper-left end of the $D_{\rm n}(4000)–L_{\mathrm{bol}}/L_{\mathrm{Edd}}$  sequence 
revealed by Wang et al. (2023) for local CL-AGNs, because of their small $D_{\rm n}(4000)$ and high 
$L_{\mathrm{bol}}/L_{\mathrm{Edd}}$, 

In summary,  the inclusion of the three CL-NLS1s shows no impact on our previous 
conclusion that 
CL-AGNs tend to be dominated by intermediate-age populations and 
to be at a specific evolutionary stage in the context of 
the coevolution of SMBHs and their host galaxies.

\section{Conclusions and Discussion}

Local CL-NLS1s are systematically searched for in this study based upon the large SDSS DR16 NLS1 sample recently compiled by Paliya et al. (2024), 
The search allows us to identify only three cases from a parent sample of 884 NLS1s ($z<0.45$) by comparing their 
SDSS DR16 and SDSS DR19 multi-epoch spectra.  
Analysis of their ``turn-off'' spectra reveals a relatively young stellar population with  $1.20<D_{\rm n}(4000) <1.45 $ associated with the three CL-NLS1s. 
The three CL-NLS1s are found to follow the previously studied local CL-AGNs in both the
H$\delta_{\mathrm{A}}-D_{\rm n}(4000)$ and $D_{\rm n}(4000)–L_{\mathrm{bol}}/L_{\mathrm{Edd}}$ sequences by occupying the small $D_{\rm n}(4000)$ and large $L_{\mathrm{bol}}/L_{\mathrm{Edd}}$ end. 
In spite of their younger stellar populations, the inclusion of the three CL-NLS1s within the broader 
CL-AGN population does not change our previous conclusion that CL-AGNs tend to be associated with intermediate-age populations. 

\subsection{Fraction of CL-NLS1s}
Our systematic search of CL-NLS1s reinforces that the CL phenomenon is quite rare in NLS1s, typically with small $M_{\mathrm{BH}}$ accreting at a 
high rate. Specifically, the fraction of CL-NLS1s is found to be only $3/884\approx0.3\%$ in the current study, 
much lower than the fraction of local CL-AGNs of $\sim5\%$ recently claimed by Wang et al. (2023) and far larger than the fraction of  
0.007\%-–0.11\% previously claimed in quasars  (e.g., Yang et al.  2018; Yu et al. 2020).  
Generally speaking, both the current search of CL-NLS1s and the search of CL-AGNs in Wang et al. (2023)
are based on comparable timescale of $\sim10-20$ yr. 
In addition to the same early SDSS DR16 database used in both studies, 
the epoch of new spectroscopy in Wang et al. (2023) ranges in MJD$\sim59000-60000$ day that is 
comparable with the period of SDSS-V, i.e., SDSS DR19 (see Figure \ref{fig:lc} in this study and 
Figure 8 in Wang et al. (2023)).

One should bear in mind that all three CL-NLS1s identified in this study 
possess relatively weak \ion{Fe}{2} complex emission, R4570 = \ion{Fe}{2}/H$\beta < 0.5$ (see Table 2). 
The implication of 
this extremely low fraction can be learned from subsequent discussion.
In addition, we propose that the CL phenomenon is difficult to identify 
in observation because of some selection effects. 
The Balmer lines of NLS1s are indeed very bright, since they are 
accreting near the Eddington rate. 
Any mild (factor of a few) weakness in the broad emission lines would still preserve 
the Type I classification of the system, not causing a well-recognizable type 
change (see also Komossa \& Grupe 2024).

\subsection{Physics of the CL Phenomenon}

\begin{figure*}[htp!]
\plotone{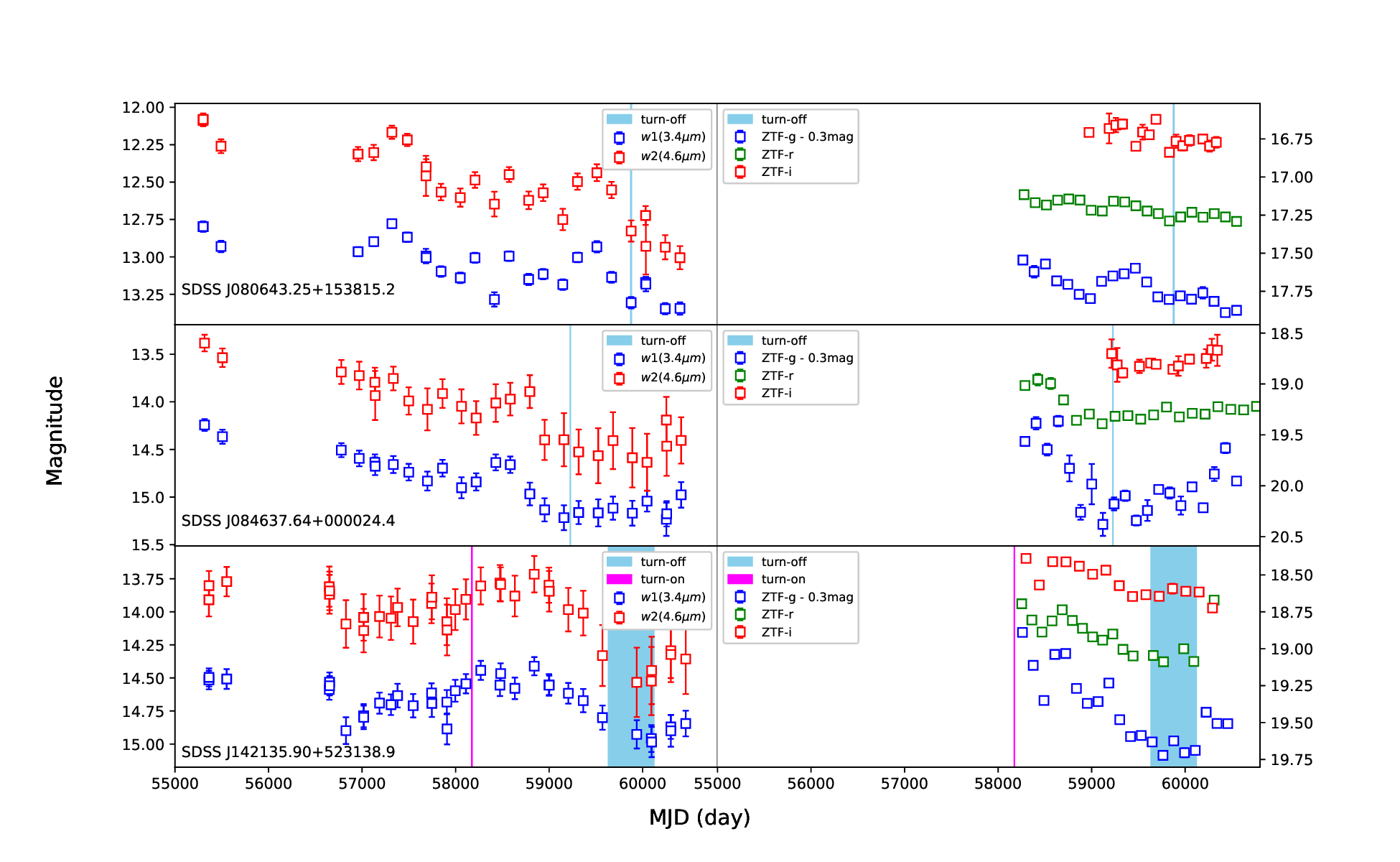}
\caption{Brightness variation of the three CL-NLS1s in MIR {\it WISE} $w1$ and $w2$ bands (the 
left panels), and in optical bands (the right panels), after binning the measured magnitudes within a couple of days. In each
panel, the vertical lines or region mark the epochs when the optical spectra were obtained.
\label{fig:lc}}
\end{figure*}

We first argue against  obscuration as the origin of the CL phenomenon observed in
the three objects. At first, as shown in Table 1, the Balmer decrements measured from the differential spectra range from 
2.7 to 3.5, consistent with standard Case B recombination. As a similar argument against the obscuration scenario,
the \ion{Mg}{2}$\lambda2800$ emission line is found to be less variable than the 
Balmer emission lines in a fraction of identified CL-AGNs (e.g., Wang et al. 2018; Zeltyn et al. 2024).  
 Secondly, by assuming the 
obscuration material orbits outside the BLR, the crossing time of 
$t_{\mathrm{cross}} = 0.11(r_{\mathrm{orb}}/{\mathrm{1\ ld}})^{3/2}(M_{\mathrm{BH}}/10^8M_\odot)^{-1/2}$ 
(LaMassa et al. 2015) is estimated to be $>25$ yr for the three CL-NLS1s
\footnote{The orbital radius of the obscuration material ($r_\mathrm{orb}$) is assumed to be equal to the size of BLR ($R_{\mathrm{BLR}}$) in the  
estimation, where the radius-luminosity relationship of 
$\log(R_{\mathrm{BLR}}/\mathrm{1\ ld})=1.559+0.549\log(\lambda L_\lambda(5100\AA)/10^{44}\ \mathrm{erg\ s^{-1}})$ 
(Bentz et al. 2013) is adopted for calculating $R_{\mathrm{BLR}}$.}, which is much larger than CL timescales 
listed in Table \ref{tab:properties}. Finally, as shown in the left panels of Figure \ref{fig:lc}, the obscuration scenario is disfavored for the three
objects by their CL-state related variations in the middle-infrared (MIR) because the MIR emission is
less sensitive to extinction. 

The left panels in 
Figure \ref{fig:lc} show the $w1$ (3.4~$\mu$m) and $w2$ (4.6~$\mu$m)  MIR light curves of the three CL-NLS1s, 
based on the {\it WISE} and {\it NEOWISE-R} surveys 
(Wright et al. 2010; Mainzer et al. 2014).
The ``turn-off'' states correspond to an MIR dimming in all three CL-NLS1s,
although the 
``turn-on'' states shown in the SDSS-DR17 dataset are not covered by the {\it WISE} survey in
both SDSS~J080643.25+153815.2 and SDSS~J084637+000024. 
Both ``turn-on'' and ``turn-off'' states are fortunately covered 
by the MIR curve in SDSS~J142135.90+523138.9, 
where the ``turn-on'' and 
``turn-off''states exactly correspond to the MIR bright and 
faint phases, respectively.

So far, there is a great deal of evidence supporting that the observed CL phenomenon results from  variations of the accretion rate onto the 
central SMBH, based on either the variation in mid-infrared (MIR) being synchronous with the CL phenomenon or low polarization,  $<1\%$ in the
``turn-off'' state (e.g.,  Sheng et al. 2017; Stern et al. 2018; Wang et al. 2019, 2020, 2021; 2023; Yang et al. 2018; Lopez-Navas et al. 2022, 2023; Hutsemekers et al. 2019).  
As an additional test, the right panels of Figure \ref{fig:lc} show
the optical light curves extracted from the ZTF survey (e.g., Shappee et al. 2014; 
Kulkarni 2018), which reinforces the accretion scenario in the three CL-NLS1s by the tight relation
between spectral state and AGN's optical brightness.

Since CL-AGNs have higher detection rate in AGNs with low $L_{\mathrm{bol}}$ and low $L_{\mathrm{bol}}/L_{\mathrm{Edd}}$\
 (e.g., MacLeod et al. 2019; Wang et al. 2023),  some authors  therefore attributed the CL phenomenon to the disk-wind broad-line region 
 (BLR) models formulated by Elitzur \& Ho (2009) and Nicastro (2000).
 By depending on a critical radius of the accretion disk where the power deposited into the
vertical outflow is maximized, the scenario proposed by Nicastro (2000) predicts a critical value of $L_{\mathrm{bol}}/L_{\mathrm{Edd}} \approx (2$--3) $\times 10^{-3}$ for the (dis)appearance of the BLR around a SMBH with a mass in a range $10^{7-8}~{\rm M}_\odot$.

In the three CL-NLS1s, this critical value is, however, lower than the $L_{\mathrm{bol}}/L_{\mathrm{Edd}}$  estimated at 
their ``turn-off'' states by at least 1--2 orders of magnitude.
This discrepancy arises because, as noted in Wang et al. (2023), the reported $L_{\mathrm{bol}}/L_{\mathrm{Edd}}$ is 
based on the weakened broad H$\alpha$ emission lines,  
and can be eliminated if the $L_{\mathrm{bol}}/L_{\mathrm{Edd}}$ is instead inferred from the X-ray emission when 
the BLR lines disappear entirely in some particular cases (e.g., Wang et al. 2020, 2022, 2023; Ai et al. 2020). 
Because of the typically high $L_{\mathrm{bol}}/L_{\mathrm{Edd}} \approx 1$ in NLS1s, the disk-wind model implies the CL phenomenon  
occurs more rarely in NLS1s than in broad-line Seyfert 1 galaxies (BLS1s) when a nearly constant critical value of $L_{\mathrm{bol}}/L_{\mathrm{Edd}}$ is adopted.  
That is to say, a much more significant decrease in accretion rate is required for the CL phenomenon occurring in NLS1s than in BLS1s. Such scenario naturally explains the occurrence of the CL phenomena in the three CL-NLS1s,
because the three objects have modest $L/L_{\mathrm{Edd}}\sim0.1$ in their 
``turn-on'' states, i.e., far below the Eddington limit. Such values are, in fact, comparable to those found in other CL-AGNs (e.g., MacLeod et al. 2019;
Wang et al. 2019; Zeltyn et al. 2024).   

We argue that the local disk thermal instability is a plausible explanation of the CL phenomenon occurring in the 
three CL-NLS1s, because the existence of a BLR requires a thermally unstable disk region dominated by radiation pressure in
the disk-wind model.
The thermal timescale of evolutionary $\alpha$-disk model  is (Siemiginowska et al. 1996)
\begin{equation}
\small
 t_{\mathrm{th}} \approx \frac{1}{\alpha\Omega_{\mathrm{K}}}\\
 =2.7\bigg(\frac{\alpha}{0.1}\bigg)^{-1}\bigg(\frac{r}{10^{16}\,\mathrm{cm}}\bigg)^{3/2}\bigg(\frac{M_{\mathrm{BH}}}{10^8~M_\odot}\bigg)^{-1/2} \mathrm{yr}\, . 
\end{equation}
The estimated $M_{\mathrm{BH}}$ yields a timescale of 8--13~yr for the three CL-NLS1s, when  the fiducial values of $\alpha=0.1$ 
(the ``viscosity parameter'' ) and $r=10^{16}~\mathrm{cm}$ are adopted in the estimation. 
A few other scenarios, alongside the thermal instability, have been proposed in literature to generate 
a shorter or comparable variability timescale. These scenarios include 
a narrow unstable zone (Sniegowska et al. 2020),
a magnetically elevated  accretion disk  (e.g., Ross et al. 2018; Stern et al. 2018; Dexter \& Begelman 2019; Pan et al. 2021),
and a thin disk that collapsed from an inner advection-dominated accretion flow (Li \& Cao 2025).

We argue that the classical viscous radial inflow has difficulty explaining the observed CL phenomenon in the three CL-NLS1s,
although the magnetic field are able to substantially shorten the inflow timescale in magnetic outflows (Feng et al. 2021b).
One can estimate the viscous timescale of a radial inflow as
(e.g., Shakura \& Sunyaev 1973; Krolik 1999; LaMassa et al. 2015; Gezari et al. 2017)

\begin{equation}
\scriptsize
 t_{\mathrm{infl}} = 6.5\bigg(\frac{\alpha}{0.1}\bigg)^{-1}\bigg(\frac{L/L_{\mathrm{Edd}}}{0.1}\bigg)^{-2}\bigg(\frac{\eta}{0.1}\bigg)^2\bigg(\frac{r}{r_g}\bigg)^{7/2}\bigg(\frac{M_{\mathrm{BH}}}{10^8~ M_\odot}\bigg)~\mathrm{yr}\, ,
\end{equation}
where $r_g$ and $\eta$ are the gravitational radius in units of $GM/c^2$ and 
the efficiency of converting potential energy to radiation, respectively. Adopting the fiducial values of $\alpha=\eta=0.1$ yields an estimate of 
$10^{5-6}$~yr for the three CL-NLS1s, when $r\approx(50$--100)\,$r_g$ is adopted in the calculation. 
This radius corresponds to the outer disk region typically generating optical emission. 

\subsection{Special Evolutionary Stage of CL-AGNs}

We have argued above that the result that CL-AGNs tend to be associated with intermediate-old stellar populations
(see also Liu et al. 2021,  Jin et al. 2022) remains valid after including the three CL-NLS1s with young associated stellar populations,
although some studies report that there is no significant difference between the host galaxies of CL-AGNs and non-CL-AGNs 
(e.g., Charlton et al. 2019; Yu et al. 2020; Dodd et al. 2021; Verrico et al. 2025).
Based on their special stellar populations,
Wang et al. (2023) proposed that CL-AGNs probably represent a transition between the ``feast'' and ``famine'' SMBH fueling stages.
The two evolutionary stages are described by Kauffmann \& Heckman (2009) by analyzing the distribution of
 $L_{\mathrm{bol}}/L_{\mathrm{Edd}}$ as a function 
of $M_{\mathrm{BH}}$ and properties of the host galaxies for a large sample of SDSS local Seyfert 2 galaxies. 
The ``feast'' phase is characterized by—and sustained through—a plentiful supply of cold gas. 
While, after the reservoir of cold gas is mainly consumed in the ``feast'' phase, the activity of SMBHs is sustained 
by slow stellar winds from evolved stars, or by accretion triggered by minor mergers or by multiple collisions of intergalactic
cold-gas clumps. 
(e.g., King \& Pringle 2006, 2007; Kauffmann \& Heckman 2009; Davies et al. 2004, 2017; Pizzolato \& Soker 2005; Gaspari et al. 2013; Heckman \& Best 2014, for a review). The two phases are separated at $D_{\rm n}(4000) = 1.5$–1.8, 
a range that roughly agrees with the average and standard deviation of $D_{\rm n}(4000)$ measured for the CL-AGNs,
i.e., $\langle D_{\rm n}(4000)\rangle = 1.51\pm0.18$ after combining the three CL-NLS1s with the 
CL-AGNs studied in Wang et al. (2023).

The special evolutionary stage of CL-AGNs is further supported by other studies. 
Dodd et al. (2021) proposed that CL-AGNs tend to have high detection rate in the galaxies with high-density pseudo-bulges.
Likely driven by AGN's feedback. these galaxies. located in the so-called ``green valley'', are believed to be undergoing a transition 
from active star-forming to quiescent elliptical systems.
Based on the $L_{\mathrm{bol}}/L_{\mathrm{Edd}}$ measured in the MIR, 
a scenario described by a transition between a Shakura-Sunyaev disk (Shakura \& Sunyaev 1973) and radiatively inefficient accretion flow
has been proposed by Lyu et al. (2022) based on the measured MIR $L_{\mathrm{bol}}/L_{\mathrm{Edd}}$. 
A possible specific evolutionary stage with $L_{\mathrm{bol}}/L_{\mathrm{Edd}}\approx0.1$ is argued for CL-AGNs by 
Liu et al. (2021), based upon a switch from a positive to a negative $\Gamma–L_{\mathrm{X}}/L_\mathrm{Edd}$ correlation in the CL phenomenon.
By focusing on a sample of repeating CL-AGNs,
Wang et al. (2024) alternatively proposed that the CL phenomenon is likely a result of a sudden change in the supply of 
circumnuclear gas during the transition between the ``feast'' and ``famine'' fueling phase. The rarity of CL-NLS1s can be alternatively
explained by this scenario because of their very long CL cycle time. Based on the values listed in Table 1, the scenario (Eq. 3 in Wang et al. (2024))
predicts a  CL cycle time of $\sim 300$~yr for CL-NLS1s.

\subsection{Narrow-line Region of NLS1s}

Although a great deal of effort has been made in past decades to explore the nature of NLS1s by examining the properties (e.g., kinematics, electron density, and ionization state) of their NLRs 
(e.g., Komossa \& Xu 2007; Xu et al. 2007; Komossa et al. 2008, 2018; Smith et al. 2008;
Berton et al. 2016; Schmidt et al. 2016, 2018; Cracco et al. 2016; Rakshit et al. 2017; 
Meena et al. 2021),  
the NLR properties of NLS1s are still an open issue at present because of the difficulty and uncertainty of 
separating narrow emission-line components from their smooth Balmer-line profiles.
Even though this difficulty could be in principle conquered by the ``turn-off'' spectra of CL-NLS1s because of the weakened broad-line emission, 
the ``fiber drop'' effect\footnote{The ``fiber drop'' effect refers to slightly misplaced fibers that miss the galaxy core, and instead point at an off-center location. 
Such off-center fibers provide spectra that miss the BLR and bright continuum 
(except for seeing-related effects), but still include the NLR.} provides us an alternative method addressing this issue. The ``fiber drop'' effect can 
mimic the CL phenomenon, and is therefore important to recognize. But it also has the advantage of providing clean NLR spectra of NLS1 galaxies.

As stated in Section 2.2, 
the ``fiber drop'' effect  has been identified in $\sim30$ cases in our cross-match 
owing to their large variation of the observed [\ion{O}{3}] $\lambda$5007 line flux,  
$\Delta f$([\ion{O}{3}])/$f_{\mathrm{DR16}} > 0.5$ (see the definition in Sec. 2.2). Visual inspection of the $\sim30$ cases shows that they can be divided into two kinds.
In one, there is an obvious reflection in their Balmer-line profiles even at their ``turn-on'' states. In the other, the 
signal-to-noise ratio of the ``fiber dropped'' spectra is too low to perform further spectral analysis in detail.

\begin{figure*}[htp!]
\plotone{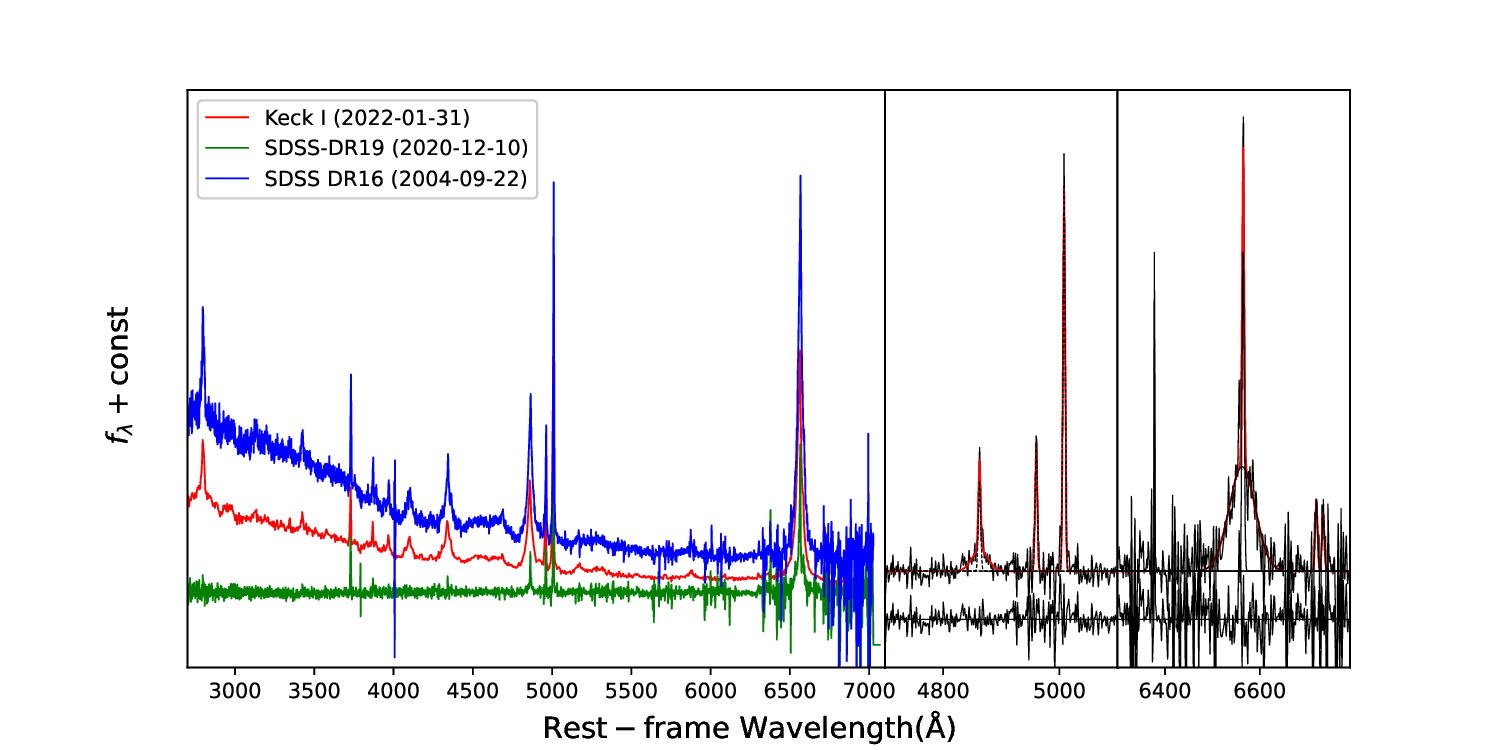}
\caption{\it Left panel: \rm a comparison of the SDSS DR16, SDSS-V DR19, and Keck spectra of NLS1 SDSS J025951.22+003744.2, after binning by a boxcar of 3\,\AA. 
{\it Middle and right panels:}  Line-profile modeling in the SDSS-V DR19  spectrum.
In each panel, the underlying continuum modeled by a local linear function determined by the continuum adjacent to the emission lines
has already been removed from the originally observed spectrum.  The line profiles in the H$\beta$ and H$\alpha$ regions are fitted by
a linear combination of a set of Gaussian functions. 
The observed and modeled line profiles are plotted with black and red solid lines, respectively. Each Gaussian function used is shown with a black dashed line. The subpanel below the line spectrum shows the residuals between the observed and modeled profiles.
\label{fig:J02595122+003744}}
\end{figure*}

This result motivates us to perform a more comprehensive search for the ``fiber drop'' effect of objects with NLS1-like spectra.   
A successful cross-match is fortunately found in SDSS J025951.22+003744.2 with $\Delta f(\mathrm{[O~III]})/f_{\mathrm{DR16}}=0.78$.
The corresponding SDSS DR16 and SDSS-V DR19 spectra are 
compared in the left panel of Figure \ref{fig:J02595122+003744}. As shown in the plot, the SDSS DR16 spectrum is typical of an NLS1 with a blue continuum, 
narrow and smooth Balmer line profiles,  strong \ion{Fe}{2} complex, small SMBH, and high accretion rate\footnote{Our spectral analysis gives 
$\mathrm{FWHM(H\beta) = 1770\pm100\ km\ s^{-1}}$, $\mathrm{R4570 = Fe~II/H\beta = 0.44}$, $M_{\mathrm{BH}} = 4.7\times10^7~M_\odot$, 
and $L_{\mathrm{bol}}/L_{\mathrm{Edd}}=0.6$; see Appendix A for details.}.
However, a global and significant dimming is found in the SDSS-V DR19 spectrum that is dominated by the narrow emission lines. 
By searching for historical archive data taken by large telescopes (e.g., VLT, GTC, Keck), spectra of the object have been
fortunately taken by the Keck~I 10~m telescope on 2022, Jan. 31, $\sim 1$~yr after the SDSS-V DR19 spectroscopy. 
The long-slit Keck spectrum, covering both nuclear and off-nuclear regions, is overplotted in the left panel of Figure \ref{fig:J02595122+003744} (see Appendix A.1 for
the data-reduction  details). One can see clearly from the plot that 
the Keck spectrum is typical of an NLS1 and consistent with the SDSS DR16 spectrum. 
This consistency reinforces the ``fiber drop''  effect in the SDSS DR19 spectrum\footnote{Examining the keywords of \tt PLUG\_RA \rm and \tt PLUG\_DEC \rm returns a
celestial distance of 2\arcsec\ between the SDSS DR16 and DR19 spectra.}, 
since the time interval between the Keck spectrum and the SDSS-V DR19 spectrum is so short 
that it is difficult to explain it by the CL phenomenon (see Eqs. (5) and (6)).

\begin{figure}[htp!]
\plotone{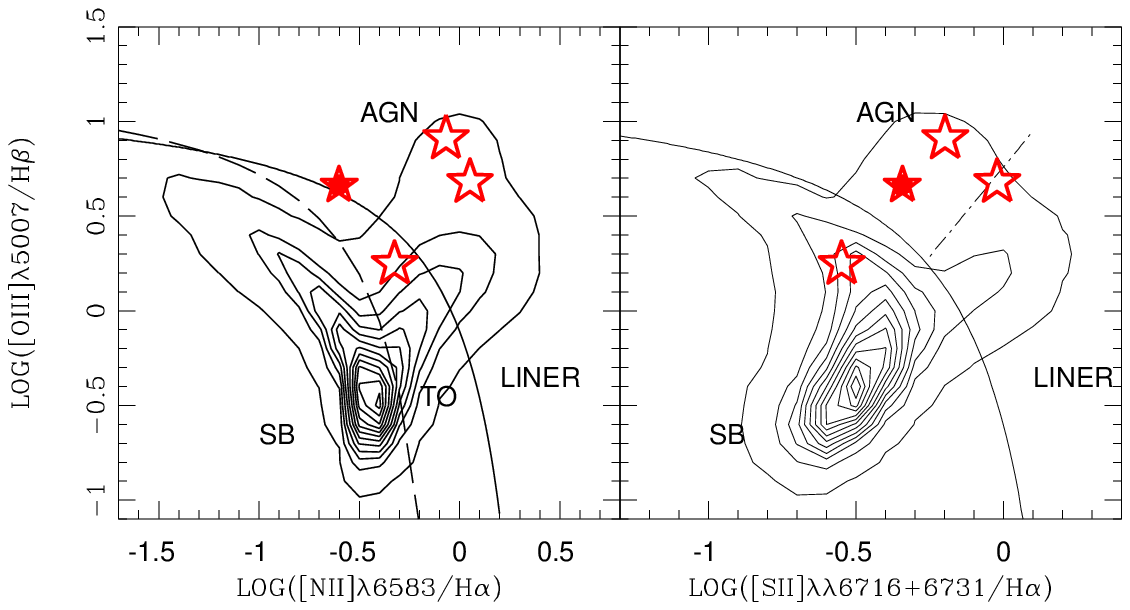}
\caption{Two of the BPT diagnostic diagrams for the NLS1 SDSS~J025951.22+003744.2 
(the solid star) based on its SDSS-V DR19 spectrum, and for the three CL-NLS1s (the open stars) based on their ``turn-off'' spectra.
The density contours
are shown for a typical distribution of the narrow-line galaxies described by Heckman et al. (2004) and Kauffmann
et al. (2003). Only the galaxies with signal-to-noise ratio $> 20$ and the emission lines detected with at least 3$\sigma$
significance are plotted. The solid lines in both panels mark the theoretical demarcation separating AGNs
from star-forming galaxies proposed by Kewley et al. (2001). The long-dashed line in the left panel shows the
empirical demarcation  proposed by Kauffmann et al. (2003), which is used to separate ``pure'' star-forming
galaxies. The dot-dashed line in the right panel is the demarcation separating Seyfert galaxies and LINERs
proposed by Kewley et al. (2006).
\label{fig:bpt}}
\end{figure}

In order to quantify the NLR properties of SDSS~J025951.22+003744.2,
the line profiles in the SDSS-V DR19 spectrum are modeled by following the method described in Section 3.2, and shown 
in the middle and right panels of Figure \ref{fig:J02595122+003744} for the H$\beta$ and H$\alpha$ regions, respectively.  With the line-profile modeling, 
the two of the \rm empirical Baldwin–Phillips–Terlevich (BPT) diagrams 
(i.e., [\ion{N}{2}]/H$\alpha$ versus [\ion{O}{3}]/H$\beta$ and [\ion{S}{2}]/H$\alpha$ versus [\ion{O}{3}]/H$\beta$)
are shown in Figure \ref{fig:bpt} for the object, along with the three CL-NLS1s
based on their ``turn-off'' spectra.
The diagrams, which were originally proposed by Baldwin et al. (1981) and then refined by Veilleux \& Osterbrock (1987), are traditionally 
used as a powerful tool to determine the dominant energy source in emission-line galaxies according to their emission-line ratios.
One can see clearly from the plots that all the NLS1s are located in the typical AGN region, except SDSS J142135.90+523132820.9. 
In both BPT diagrams, this object is located well below the widely adopted theoretical demarcation lines separating
AGNs from star-forming galaxies (Kewley et al. 2006), indicating strong circumnuclear star-formation
activity occurring in the object.

\acknowledgments

The authors thank the anonymous referee for his/her very careful review and quite helpful suggestions 
improving the manuscript. 
This study is supported by the National Natural Science Foundation of China under grants 12273054 and  12173009, and the Strategic Pioneer Program on Space Science, Chinese Academy of Sciences, grants XDA15052600 and XDA15016500. 
The authors are grateful for support from the National Key Research and Development Project of China (grant 2020YFE0202100). S.K. is partially supported by the CAS President's International Fellowship Initiative (under grant 2025PAV0180).
A.V.F.'s research group at U.C. Berkeley received financial assistance from the Christopher R. Redlich Fund, as well as donations from Gary and Cynthia Bengier, Clark and Sharon Winslow, Alan Eustace and Kathy Kwan, William Draper, Timothy and Melissa Draper, Briggs and Kathleen Wood, Sanford Robertson (W.Z. is a Bengier-Winslow-Eustace Specialist in Astronomy, T.G.B. is a Draper-Wood-Robertson Specialist in Astronomy), 
and numerous other donors.

Some of the data presented herein were obtained at Keck Observatory, which is a private 501(c)3 nonprofit organization operated as a scientific partnership among the California Institute of Technology, the University of California, and the National Aeronautics and Space Administration (NASA). The Observatory was made possible by the generous financial support of the W. M. Keck Foundation. 
The authors wish to recognize and acknowledge the very significant cultural role and reverence that the summit of Maunakea has always had within the Native Hawaiian community. We are most fortunate to have the opportunity to conduct observations from this mountain.
 
This study used the NASA/IPAC Extragalactic Database (NED), which is operated by the Jet Propulsion
Laboratory, California Institute of Technology. 
It also used data collected by the {\it Wide-field Infrared Survey Explorer (WISE)}, which is a joint project
of the University of California at Los Angeles and the Jet Propulsion Laboratory/California Institute of
Technology, funded by NASA.

\vspace{5mm}
\facilities{Keck~I 10~m telescope (LRIS)}
\software{IRAF (Tody 1986, 1992), MATPLOTLIB (Hunter 2007) }
%
\clearpage
\appendix
\section{Observation, Data Reduction, and Spectral Analysis of NLS1 SDSS~J025951.22+003744.2}

The object, SDSS~J025951.22+003744.2, was observed by the Low Resolution Imaging Spectrometer (LRIS; Oke et al. 1995) 
mounted on the Keck~I 10~m telescope on the night of 2022 January 31 (UTC). 
Using the D5600 dichroic, the spectrum was taken with the 600/4000 blue grism (dispersion 
$\mathrm{0.63\,\AA\ pixel^{-1}}$) and the 
400/8500 red grating (dispersion $\mathrm{1.20\,\AA\ pixel^{-1}}$),  covering 3150--10,270~\AA. 
The slit width was 1\arcsec\ during the observation, and the exposure time was 900~s.  We reduced the raw data 
by using the LPipe pipeline (Perley 2019), which performs a completely automated, end-to-end reduction of LRIS spectra.

The extracted one-dimensional spectrum was then calibrated in wavelength and in flux with spectra of the comparison lamps
and standard stars. The telluric A-band (7600–-7630~\AA) and B-band (around 6860~\AA) absorption produced by 
atmospheric $\mathrm{O_2}$ molecules were removed from the extracted spectra by using the standard-star spectra.
The calibrated spectrum was then corrected for Galactic extinction according to the color excess
$E(B-V)$ taken from the Schlafly \& Finkbeiner (2011) Galactic reddening map, by assuming the $R_V = 3.1$ extinction law 
of our Galaxy (Cardelli et al. 1989). The spectrum was then transformed to the rest frame according to the corresponding redshift.

With an additional component of ultraviolet (UV) \ion{Fe}{2} complex, we model the continuum of both the SDSS DR16 and Keck spectra by following the method described in Section 3.2. The UV \ion{Fe}{2} complex is modeled by the 
theoretical template by Bruhweiler \& Verner (2008).
The predicted spectrum giving the best fit to the observed I~Zw~1 spectrum is used in our fitting, which is calculated for $\log(n_{\rm H}/\mathrm{cm^{-3}})=11.0$, $\log(\Phi_{\rm H}/\mathrm{(cm^{-2}\ s^{-1}}))=20.5$, $\xi/(\mathrm{1\ km\ s^{-1}})=20$, and 
830 energy levels. In advance of the fitting, the line width of the template is fixed to be that of the broad component of H$\beta$
by convolving with a Gaussian profile whose width is determined by our line-profile modeling.

After subtracting the fitted continuum from the observed spectrum, we model the line profiles in both the H$\beta$ and H$\alpha$ regions, 
and calculate $M_{\mathrm{BH}}$ and $L_{\mathrm{bol}}/L_{\mathrm{Edd}}$ in the manner described in the main text.


\end{document}